\documentclass[a4paper,11pt]{amsart}
\usepackage{bm}
\usepackage{amsmath,amssymb}

\numberwithin{equation}{section}

\newtheorem{theorem}{Theorem}[section]
\newtheorem{lemma}[theorem]{Lemma}
\newtheorem{proposition}[theorem]{Proposition}
\newtheorem{corollary}[theorem]{Corollary}

\theoremstyle{definition}

\theoremstyle{remark}
\newtheorem{remark}[theorem]{Remark}

\begin{document}
\newcommand{\on}[1]{\mathcal{O}_{\mathbb{P}^1}(#1)}
\newcommand{\tr}{\textrm{tr}}
\newcommand{\id}{\operatorname{id}}
\newcommand{\Lie}{\operatorname{Lie}}
\newcommand{\Pic}{\operatorname{Pic}}
\newcommand{\Div}{\operatorname{Div}}
\newcommand{\rank}{\operatorname{rank}}
\newcommand{\C}{\operatorname{\mathbb{C}}}
\newcommand{\R}{\operatorname{\mathbb{R}}}
\newcommand{\Q}{\operatorname{\mathbb{Q}}}
\newcommand{\Z}{\operatorname{\mathbb{Z}}}
\newcommand{\N}{\operatorname{\mathbb{N}}}

\pagestyle{plain}


\title{\large 
Jacobian variety and integrable system\\
--- after Mumford, Beauville and Vanhaecke}
\footnote[0]{The previous version of this manuscript
was entitled 
`Integrable Hamiltonian system on 
the Jacobian of a spectral curve --- after Beauville'}


\author{Rei Inoue}
\address{Department of Physics, Graduate School of Science,
The University of Tokyo,
7-3-1 Hongo, Bunkyo, Tokyo 113-0033, Japan}
\email{reiiy@monet.phys.s.u-tokyo.ac.jp}

\author{Yukiko Konishi}
\address{Graduate School of Mathematical Sciences, The University of Tokyo,
 3-8-1 Komaba, Meguro, Tokyo 153-8914 Japan}
\email{konishi@ms.u-tokyo.ac.jp}

\author{Takao Yamazaki}
\address{Mathematical Institute, Tohoku University,
  Aoba, Sendai 980-8578, Japan}
\email{ytakao@math.tohoku.ac.jp}

\begin{abstract}
  Beauville \cite{Beauville90} introduced 
  an integrable Hamiltonian system
  whose general level set is isomorphic to 
  the complement of the theta divisor 
  in the Jacobian of the spectral curve.
  This can be regarded as a generalization of 
  the Mumford system \cite{Mumford-Book}.
  In this article, we construct a variant of Beauville's system
  whose general level set is isomorphic to
  the complement of the {\it intersection} of
  the translations of the theta divisor 
  in the Jacobian.
  A suitable subsystem of our system 
  can be regarded as a generalization of 
  the even Mumford system
  introduced by Vanhaecke \cite{Van92,Van1638}.
\end{abstract}


\renewcommand{\subjclassname}{%
  \textup{2000} Mathematics Subject Classification}

\maketitle

\section{Introduction}

The Mumford system \cite{Mumford-Book} 
is an integrable Hamiltonian system with the Lax matrix
\begin{align}
  \label{Lax-Mum}
  A(x) = 
  \begin{pmatrix}
    v(x) & w(x) \\
    u(x) & -v(x)
  \end{pmatrix}
  \in M_2( \mathbb{C}[x] ).
\end{align} 
Here 
$u(x)$ and $w(x)$ are monic of degree $d-1$ and $d$,
and  $v(x)$ is of degree $\leq d-2$
where $d$ is a fixed positive integer.
The space of Lax matrices $A(x)$ is endowed with
$d-1$ independent Hamiltonian vector fields,
defining an algebraically completely integrable dynamical system.
Its general level set is isomorphic to 
the complement of the theta divisor in the Jacobian of 
the spectral curve of the Lax matrix,
which is a hyperelliptic curve of genus $d-1$.
See \cite{AdlerMoer89, AdlerMoerVan04} for the definition of 
algebraically completely integrability.

A variant called the even Mumford system was introduced by
Vanhaecke \cite{Van92,Van1638},
whose Lax matrix has the same form as \eqref{Lax-Mum}
but the polynomial $w(x)$ is monic of degree $d+1$.
This small difference gives rise to another type of 
general level set, which is isomorphic to
the complement of the 
{\it union} of two translates of the theta divisor
in the Jacobian of a hyperelliptic curve.

On the other hand, Beauville \cite{Beauville90}
introduced a generalization of the Mumford system.
The Lax matrix is given by $A(x) \in M_r(\mathbb{C}[x])$
with a certain condition on the degree of each entry,
where $r \geq 2$ can be an arbitrary integer.
He constructed a completely integrable Hamiltonian system on
the space of (the gauge equivalence classes of) the Lax matrix $A(x)$.
Its general level set is isomorphic to the complement of 
the theta divisor in the Jacobian of the spectral curve
of the Lax matrix, which is not hyperelliptic in general.
The Mumford system can be recovered 
as the case $r=2$ of Beauville's system.

In this paper, we employ Beauville's method to construct
a system which generalize the even Mumford system.
The Lax matrix is again given by $A(x) \in M_r(\mathbb{C}[x])$
with arbitrary $r \geq 2,$
but we impose a condition, different from Beauville's,
on the degree of each entry.
(Hence the spectral curve is not hyperelliptic in general.)
We construct a completely integrable Hamiltonian system on
the space of (the gauge equivalence classes of) the Lax matrix $A(x)$.
An interesting feature of this system is that
the general level set is isomorphic to the  
complement of the {\em intersection} of  
$r$ translates of the theta divisor
(Theorem \ref{th:JM-iso} and \ref{th:M/Gr-aci}),
which is not an affine variety.
In addition, we construct a family of subsystems,
which provides an open (finite) covering of our system.
The level set of each subsystem is isomorphic to 
the complement of the {\em union} of $r$ translates of the theta
divisor in the Jacobian (Theorem \ref{th:general-eMum}).
We also construct the spaces of representatives of the subsystems,
and explicitly describe
the Hamiltonian vector fields (Proposition \ref{prop:Y-projection})
and 
the correspondence between the Lax matrix and the divisor
(Proposition \ref{prop:SoV}).
The even Mumford system can be recovered
as the case $r=2$ of a subsystem.

This paper is organized as follows:
in \S 2 we study the Jacobian of the spectral curves
for the Lax matrix.
\S 3 is devoted to the construction of Hamiltonian vector fields,
and to the proof of the integrability.
In \S 4 we introduce a family of  subsystems
and show that each of them is algebraically completely integrable.
Further we construct the spaces of representatives of the subsystems,
and study the integrable structure.
The proofs of many results in \S 2 and \S 3 are given by 
a modification of the argument of Beauville \cite{Beauville90},
nevertheless we included a rather whole proof in the present paper
for the sake of completeness,
and for the importance of Beauville's argument.

\subsection*{Remarks on related works}

The theory of algebraic integrability on a Poisson manifold
was considered by Adler and van Moerbeke
\cite{AdlerMoer89,AdlerMoerVan04}.
Integrable systems described in terms of the Lax matrix 
with the (Laurent) polynomial entries were discussed by several authors 
\cite{ReySeme94,DonagiMarkman96,AdaHarHur90,Van1638}.
In \cite{PenVan98} and \cite{AbendaFedo00},
the Mumford system was generalized to other directions.
A new treatment of the Mumford system was developed in
\cite{SmirnovNakayashiki00}.
(See also \cite{InoueYamazaki05}.)

One of the reasons that make the Mumford system (and its variants) 
interesting is a connection to
many models arising from physics,
such as the Neumann system \cite{Mumford-Book},
the Moser system \cite{AdaHarPre88},
the Toda lattice \cite{FernadesVanhae01},
the Lotka-Volterra lattice \cite{FernadesVanhae01},
and the Noumi-Yamada system \cite{InoueYamazaki05}.
We hope to find a physical model 
that realizes our system in a future study.

This subsection
is included at the suggestion of the referee.
The authors thank the referee for the advice.

\section{Jacobian of the spectral curve}
\subsection{Intersection of translations of the theta divisor}

Let $C$ be a smooth projective irreducible curve of genus $g$
(over $\mathbb{C}$).
For each integer $k,$
we write $J^k$ for the space of invertible sheaves of degree $k,$
which we regard as a principal homogeneous space
under the Jacobian $J^0$ of $C.$
We define  
the {\it theta divisor} $\Theta \subset J^{g-1}$ 
by
\begin{align*}
 \Theta =& \{ L \in J^{g-1} ~|~ H^0(C, L) \not= 0 \} \\
        =& \{ \mathcal{O}_C(E) 
    ~|~ E ~\text{is an effective divisor of degree}~ g-1 ~\}.
\end{align*}
For each point $q \in C,$ 
we write $\Theta_q$ for
the translation
$\Theta + q = 
\{ L(q) = L \otimes \mathcal{O}_C(q) ~|~ L \in \Theta \}$
of $\Theta$.
This is a divisor on $J^g.$
Let $\pi: C \to \mathbb{P}^1$ be a finite morphism of degree $r.$
We define a subvariety $J'$ of $J^{g}$ by
\begin{align*}
  J' = \{ L \in J^{g} ~|~ \pi_* L \cong O \oplus O(-1)^{\oplus r-1} ~\},
\end{align*}
where we abbreviate $\mathcal{O}_{\mathbb{P}^1}$ to $O$.
(In \cite{Beauville90},
$J'$ is denoted by $J(0, -1, \cdots, -1).$)
In this subsection, we prove the following.
\begin{proposition}\label{intertheta}
  For any point $a \in \mathbb{P}^1$ 
  unramified with respect to $\pi,$
  we have
  \[ J' = J^g \setminus (\bigcap_{q \in C} \Theta_q)
      = J^g \setminus (\bigcap_{q \in \pi^{-1}(a)} \Theta_q).
  \]
\end{proposition}

It is enough to show the following two lemmas:

\begin{lemma}\label{easyhalf}
For any point $q \in C,$
we have $J^g \setminus \Theta_q \subset J'.$
\end{lemma}
\begin{lemma}\label{mainhalf}
For any point $a \in \mathbb{P}^1$ 
unramified with respect to $\pi,$
we have
\[ J' \subset
   J^g \setminus (\bigcap_{q \in \pi^{-1}(a)} \Theta_q).
\]
\end{lemma}

We need some preliminaries to prove them.
Let $L$ be an arbitrary invertible sheaf on $C.$
We can write $\pi_* L \cong \oplus_{i=1}^r O(d_i)$
for some integers $d_1 \leq d_2 \leq  \cdots \leq d_r$
such that $\deg L = g-1 + r + \sum d_i.$ 
We have
\begin{align}
\label{h-0}
   h^0(C, L) &= h^0(\mathbb{P}^1, \pi_* L) 
     = \sum_i h^0(\mathbb{P}^1, O(d_i) ) 
     = \sum_{i \in \{j \,|\, d_j \geq 0 \}} (d_i + 1),
\\
\label{h-1}
   h^1(C, L) &= h^1(\mathbb{P}^1, \pi_* L)
     = \sum_i h^0(\mathbb{P}^1, O(-2-d_i) ) 
     = - \sum_{i \in \{j \,|\,d_j \leq -2\}} (d_i+1),
\end{align}
where we used the notation
$h^*(X, F) = \dim H^*(X, F).$
This computation, together with the Riemann-Roch theorem,
implies the following two lemmas:
\begin{lemma}\label{degreeg-1}(cf. \cite{Beauville90} 1.8)
For $L \in J^{g-1}$, the following conditions are equivalent:
\[
(1)~ L \in J^{g-1} \setminus \Theta,
\quad
(2)~ h^0(C, L)=0,
\quad
(3)~ h^1(C, L)=0,
\quad
(4)~ \pi_* L \cong O(-1)^{\oplus r}.
\]
\end{lemma}

\begin{lemma}\label{degreeg}
For $L \in J^g$, the following conditions are equivalent:
\[
(1)~ L \in J' ~~(i.e. ~\pi_* L \cong O \oplus O(-1)^{\oplus r-1}),
\quad
(2)~ h^0(C, L)=1,
\quad
(3)~ h^1(C, L)=0.
\]
\end{lemma}

\noindent
{\it Proof of Lemma \ref{easyhalf}.}
For an invertible sheaf $L$ on $C,$
we have an exact sequence
\begin{equation}\label{exact}
0 \to H^0(C, L(-q)) \to H^0(C, L) \overset{s_q}{\to} \mathbb{C} 
  \to H^1(C, L(-q)) \to H^1(C, L) \to 0
\end{equation}
deduced from
the short exact sequence
$0 \to L(-q) \to L \to \mathbb{C}_q \to 0.$
Now we assume $L \in J^g \setminus \Theta_q.$
This amounts to assuming $L(-q) \in J^{g-1} \setminus \Theta,$
and Lemma \ref{degreeg-1} shows
$h^0(C, L(-q)) = h^1(C, L(-q)) = 0.$
Then the exact sequence \eqref{exact} implies
$h^0(C, L) = 1,$
which means $L \in J'$ by Lemma \ref{degreeg}.
This completes the proof.
\qed

\noindent
{\it Proof of Lemma \ref{mainhalf}.}
We take $L \in J'.$ 
By lemma \ref{degreeg}, we have $h^0(C, L)=1.$
For $q \in C,$ 
we regard $H^0(C, L(-q))$ as a subspace of $H^0(C, L)$
by the injection appeared in eq. \eqref{exact}.

Now we assume $L \in \cap_{q \in \pi^{-1}(a)} \Theta_q.$
This amounts to assuming $L(-q) \in \Theta$ 
for any $q \in \pi^{-1}(a).$
Then Lemma \ref{degreeg-1} shows that
the inclusion
$H^0(C, L(-q)) \to H^0(C, L)$
is bijective
for any $q \in \pi^{-1}(a).$
In other words,
any non-zero global section of $L$
must have a zero at $q$ for any $q \in \pi^{-1}(a).$
Therefore 
$H^0(C, L(- \pi^* a)) = \cap_{q \in \pi^{-1}(a)} H^0(C, L(-q))$
is isomorphic to $H^0(C, L),$
and we have
$h^0(C, L(-\pi^* a)) = h^0(C, L)=1.$
However, by the projection formula 
(and the assumption $L \in J'$),
we have
\[ h^0(C, L(-\pi^* a))
  =h^0(\mathbb{P}^1, \pi_* L \otimes O(-1))
  =h^0(\mathbb{P}^1, O(-1) \oplus O(-2)^{\oplus r-1} )
  = 0.
\]
This is a contradiction, and the proof is done.
\qed

\subsection{Jacobian of the spectral curve}

We fix natural numbers $r$ and $d.$
Let us consider a polynomial of the form
\[ P(x,y)=y^r+s_1(x)y^{r-1}+\cdots+ s_r(x) \]
with $s_i(x) \in \mathbb{C}[x]$ is of degree $\leq di.$
We regard $x$ as a fixed coordinate function on $\mathbb{P}^1,$
so that the equation $P(x,y)=0$ defines a finite map
$\pi: C_P \to \mathbb{P}^1$ of degree $r,$
where $C_P$ is the {\it spectral curve} of $P.$
One can define $C_P$ to be the closure of
the affine curve defined by $P(x,y)=0$
in the Hirzebruch surface of degree $d.$
More explicitly, $C_P$ can be described
by gluing two plane affine curves defined by the polynomials 
$P(x,y)$ and $z^{dr} P(z^{-1},z^{-d} w) \in \mathbb{C}[z,w]$
by the relation $x=z^{-1}, ~y=z^{-d} w.$
The aim of this subsection is to give an explicit representation
(the matrix realization)
of the variety $J'$ considered in \S 2.1
assuming $C=C_P$ is smooth (hence irreducible).
We remark that, under this assumption,
the genus of $C_P$ is $g = \frac{1}{2}(r-1)(rd-2)$.

We introduce some notations:
\begin{align*}
&S_k(x) = \{ s(x) \in \mathbb{C}[x] ~|~ \deg s(x) \leq k \},
\\ 
&M(r,d) = 
\left\{
 A(x) \in M_r(\mathbb{C}[x]) ~\Bigg|~
\begin{array}{c}
    A(x)_{11} \in S_d(x), ~~ A(x)_{1j}\in S_{d+1}(x), \\
    A(x)_{i1}\in S_{d-1}(x), ~~ A(x)_{ij} \in S_d(x),  
\end{array}
    ~(2 \leq i,j \leq r)
\right\},
\\
&V(r,d)=\{P(x,y)=y^r+s_1(x)y^{r-1}+\cdots+ s_r(x) \in \mathbb{C}[x,y]
  ~|~ s_i(x)\in S_{di}(x)\}, 
\\
  &G_r=\Bigg\{
             g(x)=
             \begin{pmatrix}
                1  & ^t \vec{b}_1x +\, ^t \vec{b}_0\\
                0  & B
             \end{pmatrix}
            \Bigg|
            B \in GL_{r-1}(\mathbb{C}),\quad 
            \vec{b}_1,\vec{b}_0  \in\mathbb{C}^{r-1}
       \Bigg\}.
\end{align*}
In this article we denote column vectors using 
a notation such as $\vec{b}.$
We write the adjoint action of $G_r$ on $M(r,d)$ as
\begin{align}
  \label{Gr-action}
  g(A(x)) = g(x)^{-1}A(x)g(x) \text{~ for } g(x) \in G_r, ~A(x) \in M(r,d).
\end{align}
Further we introduce a map: 
\begin{align*}
 &\psi:M(r,d) \to V(r,d);~
  A(x)\mapsto \det (y\mathbb{I}_r-A(x)),
\end{align*}
and define subsets of $V(r,d)$ or $M(r,d)$ as follows:
\begin{align*}
  &M_P=\psi^{-1}(P(x,y)),
  \\
  &V_{ir}(r,d) = \{ P(x,y) \in  V(r,d) ~|~ C_P ~\text{is irreducible} \},
  \displaybreak[0]
  \\
  &V_{sm}(r,d) = \{ P(x,y) \in  V_{ir}(r,d) ~|~ C_P ~\text{is smooth}                 \},
  \\
  &M_{ir}(r,d) = \psi^{-1}(V_{ir}(r,d)),
  \\ 
  &M_{sm}(r,d) = \psi^{-1}(V_{sm}(r,d)).
\end{align*}
Then we have $V(r,d) \supset V_{ir}(r,d) \supset V_{sm}(r,d)$
and $M(r,d) \supset M_{ir}(r,d) \supset M_{sm}(r,d).$
Note that each $M_P, M_{ir}(r,d)$ and $M_{sm}(r,d)$  
is stable with respect to the action of $G_r$ \eqref{Gr-action}.
For the later use we introduce a lemma:
\begin{lemma}\label{Gr-free}
  The action \eqref{Gr-action} of $G_r$ on $M_{ir}(r,d)$ is free.
\end{lemma}
\begin{proof}
  We have to show that
  the stabilizer is trivial for all $A(x) \in M_{ir}(r,d)$. 
  Since any element of $G_r$ has an eigenvalue $1,$
  this follows from the following lemma 
  on elementary linear algebra:
\end{proof}
\begin{lemma}\label{stabilizer}
  Let $K= \mathbb{C}(x)$ be the field of rational functions over $\mathbb{C}.$
  Let $r \in \mathbb{N}$, and
  suppose $A, B \in M_r(K)$ satisfies the following conditions:
  (1) $AB=BA$, (2) $B$ is not a scalar matrix,
  (3) $B$ has an eigenvalue $b$ in $K.$
  Then $\det(y \mathbb{I}_r - A) \in K[y]$ is
  a reducible polynomial in $y.$
\end{lemma}
\begin{proof}
This follows at once by noting that
the eigenspace of $B$ with respect to $b$
is a non-trivial, proper subspace of $K^{\oplus r}$
stable under $A.$
\end{proof}
We define a projection map $\eta$: 
\begin{align}\label{projection}
  \eta : M_{ir}(r,d) \to M_{ir}(r,d) /G_r.
\end{align}

In the following, we respectively write $J_P$ and $J_P'$ 
for the variety $J$ and $J'$ defined in \S 2.1
associated to $(C_P, \pi)$.
For $k \in \mathbb{Z}$
and an invertible sheaf $L$ on $C_P,$
we use a notation 
$L(k) = L \otimes \pi^\ast O(k)$.
The main result in this subsection is the following:
\begin{theorem} (cf. \cite{Beauville90} 1.4) 
\label{th:JM-iso}
Let $P(x,y)\in V_{sm}(r,d)$,
and let $\pi: C_P \to \mathbb{P}^1$ be the finite map
defined by $x.$
Then, $M_P$ is a principal fiber bundle under $G_r,$
and the base space $M_P/G_r$ is isomorphic to $J_P'.$
\end{theorem}
\begin{proof}
The first part follows from Lemma \ref{Gr-free}.
We construct a surjective map $M_P \to J_P'$ 
and show that each fiber is 
a principal homogeneous space under $G_r.$
We remark that
a matrix $A(x) \in M(r,d)$ defines an $O$-linear map
$O \oplus O(-1)^{\oplus r-1} \to O(d) \oplus O(d-1)^{\oplus r-1}.$
(Here we consider $O(d) = O(d \cdot \infty).$)
Due to \cite{bnr} (see also \cite{Beauville90} 1.4),
the set 
\begin{equation}\label{fiber}
\{ (L, v) ~|~ L \in J_P', ~v:O \oplus O(-1)^{\oplus r-1} \cong \pi_{*}L \}
\end{equation}
is in one-to-one correspondence with $M_P$
in such a way that the diagram
\begin{equation}\label{yanda}
\begin{matrix}
& O \oplus O(-1)^{\oplus r-1} &
\overset{A(x)}{\longrightarrow}
& O(d) \oplus O(d-1)^{\oplus r-1} &
\\
&{}^{v} \downarrow & &{}^{v(d)} \downarrow &
\\
& \pi_* L &
\overset{\pi_* y}{\longrightarrow}
& \pi_* L(d)  &
\end{matrix}
\end{equation}
commutes whenever $(L,v)$ corresponds to $A(x) \in M_P.$
(Note that $A(x)$ must be in $M_P$ 
because of the relation $P(x,y)=0$ in $\mathcal{O}_C.$)
By composing this correspondence with 
the `forgetful' map $(L, v) \mapsto L,$
we obtain the desired surjection $M_P \to J_P'.$
The fiber of this map over $L \in J_P'$ is 
the set of isomorphisms
$O \oplus O(-1)^{\oplus r-1} \cong \pi_{*}L \,$
which is a principal homogeneous space under $G_r$
where the action of $g(x) \in G_r$ 
is given by $v \mapsto g(x)^{-1} \circ v \circ g(x).$
(Here we regard $g(x)$ as an automorphism on 
$O \oplus O(-1)^{\oplus r-1}$ as well as $O(d) \oplus O(d-1)^{\oplus r-1}.$)
On the set $M_P,$ this action corresponds to the conjugation.
This completes the proof.
\end{proof}

\begin{remark}\label{LandA}
Given an invertible sheaf  $L\in J_P'$, 
a corresponding matrix $A(x)\in M_P$ 
is constructed in the following way.
We have to choose an isomorphism 
$v:O \oplus O(-1)^{\oplus r-1} \to \pi_{*}L.$
This amounts to a choice of
a basis of $H^0(C_P, L(1))$ 
of the form $(f_0, f_1, \ldots, f_{r-1}, xf_0)$
with $f_0 \in H^0(C_P,L).$
The multiplication by $y$ defines elements
$yf_0 \in H^0(C_P, L(d))
= (f_0 S_d(x)) \oplus (\oplus_{j=1}^{r-1} f_j S_{d-1}(x))$
and
$yf_1, \ldots, yf_{r-1} \in H^0(C, L(d+1))
= (f_0 S_{d+1}(x)) \oplus (\oplus_{j=1}^{r-1} f_j S_{d}(x)).$
Now the matrix $A(x)$ is characterized by
\begin{equation}\notag
  y(f_0,f_1,\ldots,f_{r-1})=(f_0,f_1,\ldots,f_{r-1})A(x).
\end{equation}
In other words, 
the set $M_P$ is in one-to-one correspondence 
with the set of pairs $(L, v)$
where $L \in J_P'$ and 
$v: S_1(x) \oplus \mathbb{C}^{\oplus r-1} 
\stackrel{\cong}{\longrightarrow} H^0(C_P, L(1))$.
A matrix $A(x) \in M_P$ corresponds to $(L, v)$ iff
\begin{equation}
  \begin{matrix}
  S_1(x) \oplus \mathbb{C}^{\oplus r-1} & 
  \overset{v}{\stackrel{\cong}{\longrightarrow}} &H^0(C_P, L(1)) 
  \\
  \downarrow_{A(x)}  & & \downarrow_{y}
  \\
  S_{d+1}(x) \oplus S_{d}(x)^{\oplus r-1} & 
  \overset{v(d)}{\stackrel{\cong}{\longrightarrow}} &H^0(C_P, L(d+1)) 
\end{matrix}
\end{equation}
commutes.
\end{remark}

\subsection{Characterization of a translation of the theta divisor} 

We fix $P \in V_{sm}(r,d).$ 
Let $A(x) \in M_P,$
and let $L \in J_P'$ be the corresponding invertible sheaf.
We take $a \in \mathbb{P}^1 \setminus \{ \infty \}$ 
unramified with respect to $\pi,$
so that $\pi^{-1}(a) = \{q_1, \cdots, q_r\}$
consists of $r$ distinct points.
Then $y(q_1), \cdots, y(q_r)$ 
are the distinct eigenvalues of the matrix $A(a)$.
Let $\rho_{q_i}: \mathbb{C}^r \to \mathbb{C}$
be the projection to the eigenspace
associated with the eigenvalue $y(q_i).$
For each $i=1, \cdots, r,$
we write $s_{q_i}: H^0(C_P,L) \to \mathbb{C}$ 
for the map in the exact sequence 
\eqref{exact} applied to $q=q_i$.
In this subsection, we show the following.

\begin{proposition}
  \label{prop:r-theta}
  For each $i=1, \cdots, r,$
  the following conditions are equivalent:
\[
   (1)~ \rho_{q_i}(1, 0, \cdots, 0) \not= 0,
   \qquad
   (2)~ \mathrm{Im}(s_{q_i}) \neq {0},
   \qquad
   (3)~ L \in J'_P \setminus \Theta_{q_i}.
\]
\end{proposition}

\begin{proof}
The equivalence between $(2)$ and $(3)$
is a consequence of Lemma \ref{degreeg-1}
and the exact sequence \eqref{exact},
as is shown in the same way as Lemma \ref{easyhalf}.
We show the equivalence between $(1)$ and $(2).$
We recall that the map $s_{q_i}$ is induced by
the map $\tilde{s}_{q_i}$ 
in the following short exact sequence
of sheaves on $C_P$
\[ 0 \to L(-q_i) \longrightarrow L 
     \overset{\tilde{s}_{q_i}}{\longrightarrow} 
     \mathbb{C}_{q_i} \to 0.
\]
We then have a commutative diagram
\[
\begin{matrix}
     &\pi_* L &
     \overset{\oplus \tilde{s}_{q_i}}{\longrightarrow} 
     & \oplus_{i=1}^r \pi_* \mathbb{C}_{q_i}&
\\
     & {}^{\pi_* y} \downarrow& 
     & \downarrow^{\oplus_i y(q_i)} & 
\\
     &\pi_* L(d) &
     \overset{\oplus \tilde{s}_{q_i}(d)}{\longrightarrow} 
     & \oplus_{i=1}^r \pi_* \mathbb{C}_{q_i},&
\end{matrix}
\]
where the right vertical map is defined as
the multiplication by $y(q_i)$ on the $i$-th component.
Let $v: O \oplus O(-1)^{\oplus r-1} \cong \pi_* L$ 
be the isomorphism corresponding to $A(x).$
The pull-back of this diagram by $v$ is written as
\[
\begin{matrix}
     &O \oplus O(-1)^{\oplus r-1}&
     \overset{l_1}{\longrightarrow}
     & \mathbb{C}_a^{\oplus r}&
\\
     & \downarrow^{A(x)} & 
     & \quad \downarrow^{A(a)} & 
\\
     &O(d) \oplus O(d-1)^{\oplus r-1}&
     \overset{l_2}{\longrightarrow}
     & \mathbb{C}_a^{\oplus r},&
\end{matrix}
\]
where $l_1$ and $l_2$ are defined 
simply by the direct sum of $O(k) \to \mathbb{C}_a$
for $k \in \{0,-1,d,d-1\}$.
This means that $\pi_* \mathbb{C}_{q_i}$ maps to 
the eigenspace of $y(q_i)$ in $\mathbb{C}^r$
under the isomorphism 
$v_a: \mathbb{C}_a^r \cong \oplus_{i=1}^r \pi_* \mathbb{C}_{q_i}.$
The image of the map
$H^0(\mathbb{P}^1, O \oplus O(-1)^{\oplus r-1}) \to \mathbb{C}^{r}$
induced by $l_1$
is generated by $(1, 0, \cdots, 0).$
Therefore the image of $s_{q_i}$ is non-trivial
if and only if $\rho_{q_i}(1, 0, \cdots, 0) \not= 0.$ 
This shows the proposition.
\end{proof}

\begin{remark}
Let us consider the case $a=\infty$
(still assuming that $\pi$ is unramified at $a=\infty).$
The statement of Proposition \ref{prop:r-theta}
remains true if we replace $A(a)$ by $A(\infty),$
where the $(i,j)$-component of $A(\infty)$ is
the coefficient of the leading term of $A(x)_{ij}.$
Note that, if we set $w = y/x^d,$
then $w(q_1), \cdots, w(q_r)$ 
are the distinct eigenvalues of $A(\infty).$
\end{remark}

\section{Integrable system}
\subsection{Vector Fields}

We identify
the tangent space $T_{A(x)} M(r,d)$ at $A(x) \in M(r,d)$ with 
the affine space $M(r,d)$
and write vector fields on $M(r,d)$ 
in the matrix form.
For a positive integer $p$ 
and $a\in \mathbb{C}$,
we define a vector field $\Upsilon_a^{(p)}$ on $M(r,d)$
by  the Lax form
\begin{equation}
  \label{Upsilon-field}
  \Upsilon_a^{(p)}(A(x)):=\frac{1}{x-a}[A(a)^p,A(x)].
\end{equation}
If we let $a \in \mathbb{C}$ vary,
$\Upsilon_a^{(p)}$ can be written as 
a polynomial in $a$ of degree $pd$.
For $j= 0, \cdots, pd,$
we define a vector field $Y_j^{(p)}$ to be the
coefficient of $a^j$ in this polynomial,
viz.
\begin{equation}
  \label{vector-field}
  \Upsilon_a^{(p)}=\sum_{j=0}^{pd}a^j Y_j^{(p)}.
\end{equation}

\begin{remark}\label{rmk:Ya}
For each $a \in \mathbb{C}$,
the sets of the vector fields 
$\{ \Upsilon_a^{(p)} | 1 \leq p \leq r-1 \}$ and 
$\{ \Upsilon_a^{(p)} | 1 \leq p \}$ generate
the same vector space
by Hamilton-Cayley's formula for $A(a).$
Further for each $p \geq 1,$
the sets $\{\Upsilon_a^{(p)}|a\in\mathbb{C}\}$ and  
$\{Y_j^{(p)}| 0 \leq j\leq pd\}$ 
generate the same vector space
by Vandermond's determinant formula.
\end{remark}

\begin{lemma}%
The projection map $\eta$ \eqref{projection}
induces the equality
$\eta_* \Upsilon_a^{(p)}(A(x))=\eta_*\Upsilon_a^{(p)}(g(A(x)))$
in $T_{\eta(A(x))} (M_{ir}(r,d) /G_r)$
for all $g(x)\in G_r$ and $A(x)\in M_{ir}(r,d)$.
\end{lemma}%
\begin{proof}
A vector field $X$ on $M_{ir}(r,d)$ satisfies
$\eta_* X(A(x))=\eta_*X(g(A(x)))$ in $T_{\eta(A(x))}(M_{ir}(r,d)/G_r)$
if and only if
$X(A(x))-g_*X(A(x))$ is tangent to $G_r$-orbits 
for any $g(x)\in G_r$.
A direct calculation shows that 
$\Upsilon_a^{(p)}(A(x)) -g_* \Upsilon_a^{(p)}(A(x))$ is 
a linear combination of the vector fields of
$\textrm{Lie}\, G_r$:
\begin{equation}\label{vectorfd-Gr}
  X_E(A(x)) = [E ,A(x)],
  ~~~
  \text{for } E = E_{ij}, E_{1j}, E_{1j}^\prime~
  (2\leq i,j\leq r).
\end{equation}
Here $E_{ij}$ is given by $(E_{ij})_{kl} = \delta_{ik} \delta_{jl}$,
and $E_{1j}^\prime= x E_{1j}$. 
Thus the claim follows.
\end{proof}

\begin{corollary}
  For each $a \in \mathbb{C}, 1 \leq p\leq r-1, 0 \leq j \leq pd,$
  we have  well-defined vector fields 
  $\tilde{\Upsilon}_a^{(p)}$ and $\tilde{Y}_j^{(p)}$ 
  on $M_{ir}(r,d)/G_r$ 
  which satisfies at $[A(x)] = \eta(A(x))$ 
  \begin{equation}\notag
  \tilde{\Upsilon}_a^{(p)}([A(x)])=\eta_*\Upsilon_a^{(p)}(A(x)),
  \qquad
  \tilde{Y}_j^{(p)}([A(x)])=\eta_* Y_j^{(p)}(A(x)).
\end{equation}
\end{corollary}
We collect some properties of $\tilde{Y}_j^{(p)}.$
\begin{lemma}
  \label{prop:vector-field}
\begin{enumerate}
\item 
For each $P \in V_{ir}(r,d),$
the vector field $Y_j^{(p)}$ is tangent to $M_P$
and $\tilde{Y}_j^{(p)}$ is tangent to $M_P/G_r$.
\item 
For any $i$ and $j$,
the vector fields $Y_i^{(p)}$ and $Y_j^{(q)}$ commute.
So do $\tilde{Y}_i^{(p)}$ and $\tilde{Y}_j^{(q)}$.
\item
\label{<=g}
We have $\tilde{Y}_{pd}^{(p)}=\tilde{Y}_{pd-1}^{(p)}=0$.
The dimension of the vector space generated by 
$\tilde{Y}_j^{(p)}$ 
with $1\leq p \leq r-1,~0\leq j \leq pd-2$
is at most $g.$
\end{enumerate}
\end{lemma}
\begin{proof}
1: 
A vector field on $M(r,d)$ is equivalently 
given as a derivation on the affine ring of $M(r,d)$.
We write $t_k(x) = \tr A(x)^k$ and 
let $s_k(x)$ be 
the coefficients of $y^{r-k}$ in
$\det(y\mathbb{I}_r-A(x))$ for $1 \leq k \leq r$.
By Newton's formula, 
each $s_k(x)$ is written as a function 
in $\mathbb{Q}[t_1(x),\ldots,t_{k}(x)]$.
Since $\Upsilon_a^{(p)}$ is given by the Lax form \eqref{Upsilon-field}, 
the associated derivation satisfies
$\Upsilon_a^{(p)}(t_k(x)) = 0$.
Thus we see $\Upsilon_a^{(p)}(s_k(x)) = 0$,
and the claim follows.
\\
2: 
This is shown by a direct computation.
\\
3:  
Since
$Y_{pd}^{(p)}$ and $Y_{pd-1}^{(p)}$
are tangent to $G_r$-orbits, 
$\tilde{Y}_{pd}^{(p)}$ and 
$\tilde{Y}_{pd-1}^{(p)}$ vanish.
Therefore the space in question is generated by
$\tilde{Y}_{j}^{(p)}$ with $1\leq p\leq r-1$, $0\leq j\leq pd-2.$
The number of the members is 
$\sum_{p=1}^{r-1}(pd-1)=\frac{1}{2}(r-1)(dr-2)=g$.
\end{proof}

\subsection{Translation invariance}
We have seen that
$M_P/G_r$ is isomorphic to an open subset $J_P'$ of $J_P^g$
for $P(x,y) \in V_{sm}(r,d)$
(Theorem \ref{th:JM-iso}).
We regard the restriction of the vector fields 
$\tilde{\Upsilon}_a^{(p)}$ and $\tilde{Y}_j^{(p)}$
as vector fields on $J_P'.$
In this subsection, we show that
$\tilde{\Upsilon}_a^{(p)}|_{M_P/G_r}$ and 
$\tilde{Y}_j^{(p)}|_{M_P/G_r}$
are translation invariant under 
the action of the Jacobian $J_P^0$ on $J_P^g.$

The space of translation invariant (holomorphic) 
vector fields on $J_P$ is canonically dual to $H^0(C_P,\Omega_{C_P}^1).$
Let $C_P^0$ be the set of points $q \in C_P$
such that $\pi: C_P \to \mathbb{P}^1$ is unramified at $q$
and $\pi(q) \not = \infty.$
For $q \in C_P^0$, we write $X_q$ for the
the vector field corresponding to the linear form
$\omega \mapsto \frac{\omega}{d (x-x(q))} (q)$ on $H^0(C_P,\Omega_{C_P}^1).$
(Recall we have fixed a coordinate $x$ on $\mathbb{P}^1.$)
Equivalently, $X_q$ is characterized as follows:
the short exact sequence
$0 \to \mathcal{O}_{C_P} \to \mathcal{O}_{C_P}(q) \to T_q C_P \to 0$
induces the connecting homomorphism
\[ T_q C_P \to H^1(C_P, \mathcal{O}_{C_P}). \]
The image of the vector $\frac{\partial}{\partial(x-x(q))} \in T_q C_P$
in $H^1(C_P, \mathcal{O}_{C_P})$ corresponds to $X_q$
under the Serre duality.

\begin{remark}\label{full}
If $Q$ is an infinite subset of $C_P^0,$
the vectors $X_q ~(q \in Q)$ generate 
the full space of translation invariant vector fields.
Indeed, this is equivalent to the triviality of the cokernel of
\[\bigoplus_{q \in Q} T_q C_P \to H^1(C_P, \mathcal{O}_{C_P}), \]
which is dual to the kernel of
\[ H^0(C_P, \Omega_{C_P}^1) \to \prod_{q \in Q} T^*_q C_P; \]
but this kernel is trivial by the simple fact that
any non-zero differential form has only finitely many zeros.
\end{remark}

The main result in this subsection is the following.
\begin{theorem}(cf. \cite{Beauville90} 2.2)
  \label{th:inv-vectorf}
Let 
$a\in \mathbb{P}^1$ be a point
such that
$\pi:C_P\to \mathbb{P}^1$ is unramified over $a$,
and let $\pi^{-1}(a) = \{ q_1,\ldots, q_r \}$.
Then, for each $p \geq 1$,
the vector field 
$\tilde{\Upsilon}_a^{(p)}|_{M_P/G_r}$
coincides with
$y(q_1)^{p}X_{q_1}+\cdots+ y(q_r)^pX_{q_r}$.
\end{theorem}

\begin{proof}
Let $A(x) \in M_P.$
Then $A(a)$ has $r$ distinct eigenvalues
$y(q_1), \cdots, y(q_r).$
For each $q \in \pi^{-1}(a),$
we write $\Pi_{q} \in M_r(\mathbb{C})$ for
the projector to the eigenspace of $y(q),$
and we define a vector field $\dot{A}_q$ on $M_P$ by
\[ \dot{A}_q(A(x)) = \frac{1}{x-a} [ \Pi_{q}, A(x) ]. \] 
Since $\Upsilon_a^{(p)}|_{M_P/G_r} = 
y(q_1)^p \dot{A}_{q_1} + \cdots + y(q_r)^p \dot{A}_{q_r},$
the theorem is reduced to the following lemma.
\end{proof}

\begin{lemma}
We have $\eta_*(\Upsilon_a^{(p)})(A(x)) = X_q( \eta(A(x)) )$
for any $q \in \pi^{-1}(a), A(x) \in M_P.$
\end{lemma}

\begin{proof}
In this proof, we omit to indicate $P$
and write $C=C_P, J=J_P$ etc.
Let $C_{\epsilon}$ be the scheme 
whose underlying topological space is $C$
but with the structure sheaf 
$\mathcal{O}_C[\epsilon], \epsilon^2=0.$
For $L \in J,$
the tangent space $T_L J$ is in one-to-one correspondence
with the set of invertible sheaves on $C_{\epsilon}$,
which reduce to $L$ modulo $\epsilon.$
If $q \in C^0$ and $L \in J'$,
the vector $X_{q}(L)$ corresponding to 
the invertible sheaf $L_q^{\epsilon}$ is given by
\[ 
  H^0(U, L_q^{\epsilon}) 
= 
\left\{
s + \epsilon t 
~\Big|~ 
\begin{array}{c}
   s \in H^0(U, L),~ t \in H^0(U, L(q)), \\
   s/(x-a)+t ~\text{is holomorphic at}~q
\end{array}
\right\}
\]
for an open set $U$ of $C$
(cf. \cite{Beauville90} 2.2).

Recall that the set $M_P$ is in one-to-one correspondence
with the set of pairs $(L, v)$
where $L \in J'$ and 
$v$ is an isomorphism 
$H^0(C, L(1)) \overset{\cong}{\longrightarrow} 
S_1(x) \oplus \mathbb{C}^{\oplus r-1}$
(cf. Remark \ref{LandA}).
If $A(x) \in M_P$ corresponds to $(L, v),$
the tangent space $T_{A(x)} M_P$ 
is in one-to-one correspondence with
the pairs of $(L^{\epsilon}, v^{\epsilon})$
where $L^{\epsilon}$ is an invertible sheaf on $C_{\epsilon}$
which reduces to $L$ modulo $\epsilon,$
and $v^{\epsilon}$ is an isomorphism
$(S_1(x) \oplus \mathbb{C}^{\oplus r-1}) \otimes \mathbb{C}[\epsilon]
\cong 
H^0(C_{\epsilon}, L^{\epsilon}(1))$
of $\mathbb{C}[\epsilon]$-modules,
which reduces to $v$ modulo $\epsilon.$
A vector $\dot{A}(x) \in T_{A(x)} M_P \subset T_{A(x)} 
M_{sm}(r,d) \cong M(r,d)$
corresponds to a pair $(L^{\epsilon}, v_{\epsilon})$
iff
\begin{equation}
\begin{matrix}
(S_1^{\epsilon} \oplus \mathbb{C}[\epsilon]^{\oplus r-1})
 & \overset{v^{\epsilon}}{\stackrel{\cong}{\longrightarrow}} 
  &H^0(C_{\epsilon}, L^{\epsilon}(1)) 
\\
{}_{A(x)+\epsilon \dot{A}(x)} \downarrow & & \downarrow_{y}
\\
(S_{d+1}^{\epsilon} \oplus S_{d}^{\epsilon \,\oplus r-1})
 & \overset{v^{\epsilon}(d)}{\stackrel{\cong}{\longrightarrow}} 
&H^0(C_{\epsilon}, L^{\epsilon}(d+1)) 
\end{matrix}
\end{equation}
commutes.
Here we denote $S_k^{\epsilon} = S_k(x) \otimes \mathbb{C}[\epsilon].$

Now let $q \in C^0.$
Let $A(x) \in M_P$ and let $(L, v)$ be the corresponding pair.
Recall that $L_{q}^{\epsilon}$ is the invertible sheaf on $C_{\epsilon}$
corresponding to $X_{q}(L).$
In order to complete the proof,
we are going to construct an isomorphism
$v_q^{\epsilon}:
S_1^{\epsilon} \oplus \mathbb{C}[\epsilon]^{\oplus r-1}
\overset{\cong}{\longrightarrow} 
H^0(C_{\epsilon}, L_{q}^{\epsilon}(1))$
such that 
$v_q^{\epsilon}$ reduces to $v$ modulo $\epsilon$,
and that the diagram
\begin{equation}\label{LandAdiag2}
\begin{matrix}
(S_1^{\epsilon} \oplus \mathbb{C}[\epsilon]^{\oplus r-1})
 & \overset{v_q^{\epsilon}}{\stackrel{\cong}{\longrightarrow}}
 &H^0(C_{\epsilon}, L_q^{\epsilon}(1)) 
\\
{}_{A(x)+\epsilon \dot{A}_q(x)} \downarrow & & \downarrow_{y}
\\
(S_{d+1}^{\epsilon} \oplus S_{d}^{\epsilon \, \oplus r-1})
 & \overset{v_q^{\epsilon}(d)}{\stackrel{\cong}{\longrightarrow}} 
&H^0(C_{\epsilon}, L_q^{\epsilon}(d+1)) 
\end{matrix}
\end{equation}
commutes.

Let $a = \pi(q)$ 
and write $\pi^{-1}(a) = \{ q_1 = q, q_2, \cdots, q_r \}.$
There exists a section $s_i \in H^0(C, L(1))$
which does not vanish at $q_i$ but vanish at $q_j$ 
for $j \not= i.$
However, such an $s_i$ is not unique.
We specify a choice of $s_i$ as follows.
We write $f_0, f_1, \cdots, f_{r-1} \in H^0(C, L(1))$
for the images of 
$(1, (0, \cdots, 0)), (0, (1, 0, \cdots, 0)), 
\cdots, (0, (0, \cdots, 1))$
under the isomorphism $v.$
Then
$((x-a)f_0, f_0, f_1, \cdots, f_{r-1})$
is a $\mathbb{C}$-basis of $H^0(C, L(1))$
(and $(x-a)f_0$ is 
a $\mathbb{C}$-base of $H^0(C, L)$).
On the other hand, 
$((x-a)f_0, s_1, \cdots, s_r)$ is
also a basis of $H^0(C, L(1)).$
Thus we can write
\begin{align*}
((x-a)f_0, s_1,\ldots,s_r)=((x-a)f_0, f_0, f_1, \cdots, f_{r-1}) 
\cdot \tilde{\Lambda},
\\
\tilde{\Lambda} = 
\begin{pmatrix} 1 & * \\ 0 & \Lambda \end{pmatrix},
\quad
\Lambda
=
\begin{pmatrix}
\vec{\lambda}_1,\ldots,\vec{\lambda}_r
\end{pmatrix}
\in 
GL_{r}(\mathbb{C}).
\end{align*}
We can choose $s_1, \cdots, s_r$ so that
$\tilde{\Lambda} = 
\begin{pmatrix} 1 & 0 \\ 0 & \Lambda \end{pmatrix}$.
This condition determines $s_i$ 
up to a multiplication by a non-zero scalar.
By definition we have
$((x-a)f_0/s_i)(q_i) = 0$ and $s_j/s_i(q_i) = \delta_{i,j}.$
Hence, if we set 
$\mathbf{f}:=\big( (f_j/s_i)(q_i) \big)_{ij},$
then $\mathbf{f} \cdot\Lambda=\mathbb{I}_r$.

Now we define $v_q^{\epsilon}$ to be the composition of
\begin{align}
  \label{sec}
  \begin{split}
   \sigma:~ &H^0(C, L(1)) \oplus H^0(C, L(1)) \epsilon
   \overset{\cong}{\longrightarrow} 
   H^0(C_{\epsilon}, L_q^{\epsilon}(1))
\\
  &(t_1, t_2 \epsilon) \mapsto 
   t_1 + \Bigl(t_2 - \frac{t_1}{s_1}(q) \frac{s_1}{x-a}\Bigr)\epsilon
  \end{split}
\end{align}
with an isomorphism
\[ v \otimes id_{\mathbb{C}[\epsilon]}:
 (S^{\epsilon}_1 \oplus \mathbb{C}[\epsilon]^{r-1}) 
 \overset{\cong}{\longrightarrow} 
  H^0(C, L(1)) \otimes \mathbb{C}[\epsilon]
 = H^0(C, L(1)) \oplus H^0(C, L(1)) \epsilon.
\]
The change of $s_1$ by a scalar multiplication
does not affect the definition of this map.

It is immediate that $v_q^{\epsilon} \mod \epsilon$ is $v_q.$
We check the commutativity of \eqref{LandAdiag2}.
We write
$\vec{f} = (f_0, \cdots, f_{r-1})$
and 
$\vec{f}/s_i(q)  = (f_0/s_i(q), \cdots, f_{r-1}/s_i(q)).$
Then the map \eqref{sec} can be written 
in terms of matrices
\[ \sigma(\vec{f}, \epsilon \dot{\vec{f}}) =
   \vec{f} 
   + \epsilon\Bigl(\dot{\vec{f}} - \frac{1}{x-a} \vec{f} \cdot \Pi\Bigr),
\quad 
   \Pi = \vec{\lambda}_1 \cdot \vec{f}/s_1(q) \in M_r(\mathbb{C}).
\]
Therefore, the commutativity of \eqref{LandAdiag2} means
\[ \vec{f} A(x) \Bigl(\mathbb{I} - \frac{\epsilon}{x-a} \Pi\Bigr)
   =
   \vec{f} \Bigl(\mathbb{I} - \frac{\epsilon}{x-a} \Pi\Bigr) 
   (A(x) + \epsilon \dot{A}_q(x)),
\]
which follows if we have $\Pi = \Pi_{q_1}.$
To show the last assertion,
we note that the equation
$y s_i = \vec{f} A(x) \vec{\lambda}_i$
holds in $H^0(C, L(d+1)).$
Thus we have 
$\mathbf{f} A(a) \Lambda = diag(y(q_1), \cdots, y(q_r)).$
Since $\mathbf{f} = \Lambda^{-1},$
this means $\vec{\lambda}_i$ is an eigenvector
of $A(a)$ belonging to the eigenvalue $y(q_i).$
In particular, 
$\Pi = \vec{\lambda}_1 \cdot \vec{f}/s_1(q_1)$
is the projector $\Pi_{q_1}.$
This completes the proof.
\end{proof}

By Lemma \ref{prop:vector-field}-\ref{<=g} 
and Remark \ref{full}, we obtain
\begin{corollary}
  The space of vector fields on $M_{ir}(r,d)/G_r$ generated by
  $\tilde{Y}_j^{(p)}$ 
  ($1\leq p\leq r-1,0\leq j\leq pd-2$) is $g$-dimensional.
\end{corollary}

\subsection{Hamiltonian structure}

In this subsection, we show that 
the vector fields $\tilde\Upsilon_a^{(p)}$ on $M_{ir}(r,d) / G_r$ 
are Hamiltonian,
following the method of \cite{Beauville90} \S 5
(see also \cite{Kirillov-Book} \S 15, \cite{OlshPere1994}).

Let $a_1, \ldots, a_{d+2}$ be distinct points in $\mathbb{C}$,
and $\varphi: M(r,d) \to M_r(\mathbb{C})^{d+2}$ be a map
defined by
\begin{equation}\label{defvarphi}
  \varphi(A(x))=(c_1A(a_1),\ldots,c_{d+2}A(a_{d+2})).
\end{equation}
Here $c_{\alpha}=P_{\alpha}(a_{\alpha})^{-1}$ with 
$P_{\alpha}(x)=\prod_{\rho\neq \alpha}(x-a_{\rho})$.
This map is injective, and the preimage of  
$\mathbf{Y}=(Y_1, Y_2, \ldots, Y_{d+2}) \in \varphi(M(r,d))$
is obtained as $\varphi^{-1}(\mathbf{Y}) = 
\sum_{\alpha=1}^{d+2} Y_{\alpha}P_{\alpha}(x)$ 
by Lagrange's interpolation formula.

We set the coordinate on $M_r(\mathbb{C})^{d+2}$
by using $y_{ij}^{\alpha} ~(1\leq \alpha\leq d+2,1\leq i,j\leq r)$ 
as $Y_{\alpha}=(y_{ij}^{\alpha})_{1\leq i,j\leq r} \in M_r(\mathbb{C})$
and $\mathbf{Y}=(Y_1, Y_2, \ldots, Y_{d+2}) \in M_r(\mathbb{C})^{d+2}$.
We define the $G_r$-action on $M_r(\mathbb{C})^{d+2}$ by
\begin{equation}\label{graction}
  g(x):(Y_\alpha)_{1\leq \alpha\leq d+2} 
   \mapsto
  \big(g(a_{\alpha})^{-1}Y_{\alpha}g(a_{\alpha})
  \big)_{1\leq \alpha\leq d+2},
\end{equation}
which is compatible with the $G_r$-action on $M(r,d)$.
We equip $M_r(\mathbb{C})^{d+2}$ 
with the Poisson bracket which comes from that of
$gl_r(\mathbb{C})\cong M_r(\mathbb{C})$:
\begin{equation}\label{poissonstr}
  \{y_{ij}^{\alpha},y_{kl}^{\beta}\}=\delta_{\alpha,\beta}
  (\delta_{j,k}y_{il}^{\alpha}-\delta_{l,i}y_{kj}^{\alpha}).
\end{equation}
The associated Casimir functions 
are $t_{k,\alpha} = \tr (Y_\alpha^k)$ for 
$1 \leq \alpha \leq d+2,~ k \in \mathbb{Z}_{>0}$.

For $E \in$ Lie $G_r,$
we introduce the Hamiltonian functions 
$H_E$ on $M_r(\mathbb{C})^{d+2}$: 
\begin{align*}
  &H_{E_{1j}} = \sum_{\alpha} y^\alpha_{j1}, 
  ~~~
  H_{E_{1j}^\prime} = \sum_{\alpha} a_\alpha y^\alpha_{j1},
  ~~~
  H_{E_{ij}} = \sum_{\alpha} y^\alpha_{ji},
  ~~~
  \text{ for } 2 \leq i,j \leq r.
\end{align*}
These satisfy $H_{[E,E']} = \{ H_{E},H_{E'} \}$
for any $E, E' \in$ Lie $G_r$.
Each $H_E$ generates a vector field on $M_r(\mathbb{C})^{d+2}$
compatible with $X_E$ \eqref{vectorfd-Gr} on $M(r,d)$
via the map $\varphi.$
The associated moment map 
$\mu : M_r(\mathbb{C})^{d+2} \to (\text{Lie}~G_r)^\ast$
is the unique map which satisfies
$H_E(\mathbf{Y}) = \langle \mu(\mathbf{Y}), E \rangle$ for all 
$\mathbf{Y} \in M_r(\mathbb{C})^{d+2}$ and  $E \in$ Lie $G_r$.
Here $\langle ~,~ \rangle$ is the pairing between 
$(\text{Lie}~G_r)^\ast$ and $\Lie G_r$.
 
\begin{lemma}
  \label{prop:Poisson}
  \begin{enumerate}
  \item
  The image of $\varphi$ is an affine subvariety of $M_r(\mathbb{C})^{d+2}$
  determined as the intersection of $\mu^{-1}(0)$ and $t_1^{-1}(0)$,
  where $t_1 = \sum_{\alpha} t_{1,\alpha}$.
  \item
  The Poisson structure \eqref{poissonstr}
  induces the Poisson structure on $\varphi(M_{ir}(r,d))/G_r$, and hence on 
  $M_{ir}(r,d)/G_r$ via $\varphi$.
  \end{enumerate}
\end{lemma}
\begin{proof}
  1:
  The image $\varphi(M(r,d))$ of $\varphi$
  is a subvariety of $M_r(\C)^{d+2}$ 
  determined by the following conditions:
\begin{equation}\label{imvarphi}
  \begin{split}
  &\sum_{\alpha=1}^{d+2}y_{11}^{\alpha}=0,
  \\
  &\sum_{\alpha=1}^{d+2}y_{j1}^{\alpha}=0,
  ~~~~~~
  \sum_{\alpha=1}^{d+2}a_{\alpha}y_{j1}^{\alpha}=0,
  ~~~~~~
  \sum_{\alpha=1}^{d+2}y_{ji}^{\alpha}=0,
  ~~~~~~
  \text{for} ~~2\leq i,j\leq r.
  \end{split}
\end{equation}
  We see that
  the last three conditions are nothing but the  
  defining equations for $\mu^{-1}(0)$ (i.e. the zero of the 
  Hamiltonian functions $H_E$).
  Summing up the first one and the last one for 
  $2 \leq i=j \leq r$, we obtain the defining equation for 
  $t_1^{-1}(0)$.
  \\
  2:
  Recall that the action of $G_r$ on $\varphi(M_{ir}(r,d)) 
  \subset M_r(\mathbb{C})^{d+2}_0$ is free,
  and that $\varphi(M_{ir}(r,d)) \subset \mu^{-1}(0) \cap t_1^{-1}(0)$.  
  Then the Poisson structure 
  \eqref{poissonstr} on $M_r(\mathbb{C})^{d+2}$ induces 
  the Poisson structure on the quotient space 
  $\varphi(M_{ir}(r,d)) / G_r$.
  This is passed to
  the Poisson structure on $M_{ir}(r,d) / G_r$ by $\varphi$.
\end{proof}
The following lemma is shown by a direct computation.
\begin{lemma}\label{lemma:Hamiltonian-field}
  The vector fields $(p+1)\prod_{\alpha=1}^{d+2} (a-a_\alpha) 
  \tilde{\Upsilon}_a^{(p)}$ 
  on $M_{ir}(r,d) / G_r$ is Hamiltonian.
  They are generated by the $G_r$-invariant function $\tr A(a)^{p+1}$
  on $M_{ir}(r,d)$
  with respect to the Poisson bracket of Lemma \ref{prop:Poisson}-2.
\end{lemma}

Summarizing Theorem \ref{th:JM-iso}, \ref{th:inv-vectorf}
and Lemma \ref{lemma:Hamiltonian-field},
we conclude that 
\begin{theorem}(cf. [1] 5.3)
  \label{th:M/Gr-aci}    
  The Hamiltonian system $\psi |_{M_{ir}(r,d)/G_r}
  : M_{ir}(r,d)/G_r \to V(r,d)$
  is completely integrable. 
  In particular, the general level set is isomorphic to an open subvariety 
  of a Jacobian.
  More precisely, we have $M_P/G_r \cong J'_P$
  if $P \in V_{sm}(r,d)$.
\end{theorem}

\section{Generalization of Even Mumford System}
\subsection{Matrix realization of the affine Jacobian}

In this section, 
we construct a family of subsystems of $M_{ir}(r,d) / G_r$
whose general level set is 
isomorphic to the complement of
the union of 
$r$ translates of the theta divisor in the Jacobian.

In the following, we write $A(x) \in M(r,d)$ as
\begin{equation}
  \label{A-rep}
  A(x)=
  \begin{pmatrix}
  v(x) & ^t \vec{w}(x)\\ 
  \vec{u}(x) & T(x)
  \end{pmatrix},
\end{equation}
where $v(x) \in S_d(x)$, $\vec{u}(x) \in S_{d-1}(x)^{\oplus r-1}$,
$\vec{w}(x) \in S_{d+1}(x)^{\oplus r-1}$
and $T(x) \in M_{r-1}(S_d(x))$.
The coefficients of $x^k$ ($k \geq 0$) in $v(x),\vec{w}(x),
\vec{u}(x)$ and $T(x)$ will be denoted by
$v_k,\vec{w}_k,\vec{u}_k$ and $T_k$.
For $A(x)\in M(r,d)$, we define 
\begin{align}
  \label{D-function}
  D(A(x);x) 
  &=
  \big(\vec{u}(x),T(x)\vec{u}(x),\ldots, T(x)^{r-2}\vec{u}(x)\big)
  \in M_{r-1}(\mathbb{C}[x]),
\\
  D(A(x);\infty) 
  &=
  \big(\vec{u}_{d-1},T_d\vec{u}_{d-1},\ldots, 
            {T_{d}}^{r-2}\vec{u}_{d-1}\big)
  \in M_{r-1}(\mathbb{C}).
\end{align}
Note that $\det D(A(x);x)$ is a polynomial in $x$ of degree at most
$g$, and that the coefficients of $x^g$ is
$\det D(A(x);\infty)$.

For each $c\in\mathbb{P}^1$,
we define the subspaces ${\mathcal M}_c$, ${\mathcal M}_c^{ir}$
and ${\mathcal M}_{c,P}$    
of $M(r,d)$:
\begin{align*}
    &{\mathcal M}_{c}
    =
    \{A(x)\in M(r,d)~|~ \det D(A(x);c)\neq 0 \},
    \\
    &{\mathcal M}_{c}^{ir} = \mathcal{M}_c \cap M_{ir}(r,d),
    \\
    &{\mathcal M}_{c,P} = \mathcal{M}_c \cap M_P.
\end{align*}

\begin{lemma}\label{lemma:Mc-free}
  \begin{enumerate}
  \item 
  The subset $\mathcal{M}_c$ is invariant
  under the action of $G_r$ on $M(r,d).$
  \item
  The action of $G_r$ on $\mathcal{M}_c$ is free.
  \item
  Let $c_1,\ldots,c_{g+1}$ be distinct points on $\mathbb{P}^1.$
  Then we have
  \begin{equation}\notag
    M_{ir}(r,d)
  \subset 
    \bigcup_{i=1}^{g+1} \mathcal{M}_{c_i}
  = \bigcup_{c\in\mathbb{P}^1}\mathcal{M}_c 
  \subset 
    M(r,d). 
  \end{equation}
  \end{enumerate}
\end{lemma}
\begin{proof}
  Let $A(x) \in \mathcal{M}_c$ and    
   $\displaystyle{g(x) =
   \begin{pmatrix}
    1 & ^t \vec{b}(x) \\
    \vec{0} & B 
  \end{pmatrix} \in G_r
    }$.
  \\
  1: This follows from the relation $\det D(g(A(x));x)=
     \det B^{-1} \cdot \det D(A(x);x).$
  \\
  2: A computation
\begin{equation}\label{gA}
g(A(x))
=\begin{pmatrix}
 v-{}^t\Vec{b}\cdot B^{-1} \Vec{u}&
 {}^t\Vec{w}\cdot B +v{}^t\Vec{b}
 -{}^t\Vec{b}B^{-1}\Vec{u}{}^t\Vec{b}-{}^t\Vec{b}B^{-1}TB
\\
 B^{-1}\Vec{u}&
 B^{-1}\Vec{u}{}^t\Vec{b}+B^{-1}TB
\end{pmatrix}
\end{equation}
  shows that the condition $g(A(x)) = A(x)$ implies 
  $B D(A(x); x) = D(A(x); x)$ and ${}^t \vec{b} D(A(x); x) =0.$
  If we further assume $A(x) \in \mathcal{M}_c$, 
  then we obtain $B=\mathbb{I}_{r-1}$ and $\vec{b}=0.$
  \\
  3: The equality in the middle holds since $\deg_x D(A(x);x) \leq g.$
  We show the left inclusion.
  Assume $A(x) \notin \mathcal{M}_c$ for all $c \in \mathbb{P}^1$.
  Then $D(A(x);x)$ is identically zero. 
  Hence we have
  \begin{equation}\notag
    \det\Bigg(
    \begin{pmatrix}1\\ \vec{0}\end{pmatrix},
    A(x)
    \begin{pmatrix}1\\ \vec{0}\end{pmatrix},
    \ldots,
    A(x)^{r-1}
    \begin{pmatrix}1\\ \vec{0}\end{pmatrix}
    \Bigg)
    =0,
  \end{equation}
  which implies that 
  the column vectors span a proper subspace in $\C(x)^{\oplus r}$
  invariant under $A(x)$.
  Therefore the characteristic polynomial of $A(x)$ is reducible
  if $A(x) \notin \mathcal{M}_c$.
\end{proof}

This lemma implies that 
$\mathcal{M}_c^{ir}/G_r$ 
is a subsystem of the completely integrable system $M_{ir}(r,d)/G_r$. 
The general level set is described in the following:
\begin{proposition}
  \label{prop:S-J''}
  Let $c\in\mathbb{P}^1$ and
  $P \in V_{sm}(r,d)$ such that
  $\pi:C_P\to \mathbb{P}^1$ is unramified over $c$.
  Then the level set $\mathcal{M}_{c,P} / G_r$ of 
  $\mathcal{M}_c^{ir}/G_r$ is isomorphic to 
  $J_P \setminus \bigl( \bigcup_{q \in \pi^{-1}(c)} \Theta_q \bigr)$.
\end{proposition}
\begin{proof}
Let $A(x) \in M_P$ 
and let $L \in J_P'$ be the image of $A(x)$
under the map
$M_P \to M_P/G_r \cong J_P'$.
According to Proposition \ref{prop:r-theta} and Theorem \ref{th:JM-iso},
$L$ is in 
$\cup_{q \in \pi^{-1}(c)} \Theta_q$
if and only if
the first entry of any eigenvector of 
${}^t A(c)$ is nonzero.
Thus the following lemma on linear algebra completes the proof.
\end{proof}
\begin{lemma}\label{lem:zero-entry}
Let $C \in M_r(\C)$ be a semi-simple matrix.
Writing ${}^t C = \begin{pmatrix} * & * \\ \vec{c}& C_0 \end{pmatrix}$
with $C_0 \in M_{r-1}(\C)$ and $\vec{c} \in \C^{r-1},$
we set
$D={}^t(\vec{c}, C_0 \vec{c}, \cdots,{C_0}^{r-2}\vec{c}) \in M_{r-1}(\C).$
We write $W$ for the subspace of $\C^r$
generated by all eigenvectors of $C$
whose first entries are zero.
Then we have $\dim W = r-1- \rank D.$
\end{lemma}   
\begin{proof}
Define $i: \C^{r-1} \to \C^r$ 
by setting the first entry to be zero,
and let $V_0 = i(\C^{r-1}).$
Let $W_0 = \{ i(\vec{w}) \in V_0 ~|~ \vec{w} \in \C^{r-1},~ D \vec{w}= 0 \}.$
Since $\dim W_0 = r-1- \rank D,$
it is enough to show $W=W_0.$
The lemma below shows that
$W$ is the maximal subspace of $V_0$
which satisfies the condition $C W \subset W.$
Since $CW_0 \subset W_0,$ we have $W_0 \subset W.$ 
To show the converse, we take $\vec{w} \in W.$
Since $CW \subset W$,
we have $C^k \vec{w} \in W (\subset V_0)$ for all $k \geq 0.$
By writing down the condition $C^k \vec{w} \in V_0$
for $k=0, 1, \cdots,$
we see $\vec{w} \in W_0.$
This shows $W \subset W_0$ and we have done.
\end{proof}

\begin{lemma}\label{subsublemma}
Let $f: V \to V$ be a semi-simple endomorphism of 
a finite dimensional $\C$-vector space.
For a subspace $V'$ of $V,$
we write $Ev(V')$ for the set of eigenvectors
of $f$ in $V'.$
Let $W$ be a subspace of $V.$
Let $W_{st}$ be the maximal subspace in $W$
which satisfies $f(W_{st}) \subset W_{st},$
and let $W_{eig}$ be the subspace of $V$ generated by $Ev(W).$
Then we have $W_{st} = W_{eig}.$
\end{lemma}
\begin{proof}
We have $W_{eig} \subset W_{st}$ because $f(W_{eig}) \subset W_{eig}.$
It holds that
\[ W_{st} \overset{(1)}{=} \langle Ev(W_{st}) \rangle 
          \overset{(2)}{\subset}  \langle Ev(W) \rangle 
          \overset{(3)}{=} W_{eig}. \]
Here $(1), (2)$ and $(3)$ follows 
by the semi-simplicity of $f$,
by $Ev(W_{st}) \subset Ev(W)$ 
and by definition, respectively.
\end{proof}

We summarize our main result.
\begin{theorem}
  \label{th:general-eMum}
  The Hamiltonian system 
  $\psi|_{\mathcal{M}_c^{ir} /G_r}: \mathcal{M}_c^{ir} /G_r \to V(r,d)$ 
  is algebraically completely integrable.
  In particular the general level set 
  is isomorphic to an affine subvariety of a Jacobian. 
  More precisely, if $P \in V_{sm}(r,d)$ and if 
  $\pi: C_P\to {\mathbb{P}^1}$ is unramified over $c$,
  we have $\mathcal M_{c,P}/G_r \cong 
  J_P^g \setminus (\bigcup_{q \in \pi^{-1}(c)} \Theta_q)$.
\end{theorem}

\begin{remark}
  The Hamiltonian vector fields $\tilde{\Upsilon}_a^{(p)}$
  are defined on $\mathcal{M}_c /G_r$ 
  (not only on $\mathcal{M}_c^{ir} /G_r$)
  because of Lemma \ref{lemma:Mc-free}-2.
\end{remark}

\subsection{Space of Representatives}

We introduce a space of representatives of $\mathcal{M}_c /G_r$.
For Beauville's system,
Donagi and Markman \cite{DonagiMarkman96}
constructed such a space of representatives.

We define subspaces 
$\mathcal{S}_c$ of $M(r,d)$ for $c\in \mathbb{P}^1$ as follows:
\begin{align*}
    {\mathcal S}_c &= 
    \Bigg
    \{A(x) \in M(r,d) \Bigg|
    A(x)=
    \begin{pmatrix}
      v^{(0)} & {}^t\vec{w}^{(0)} \\
      \vec{\nu} & \tau
    \end{pmatrix}
    +
    (x-c)
    \begin{pmatrix}
      v^{(1)} & {}^t\vec{w}^{(1)} \\
      \vec{u}^{(1)}& T^{(1)}
    \end{pmatrix}
    \\
    & \hspace{4cm} 
    +\text{ higher terms in $(x-c)$},
    ~~~~  T^{(1)} \in \mathcal{T}
    \Bigg\},     
    \qquad \text{for $c\in\mathbb{C}$},
    \displaybreak[0]\\
   {\mathcal S}_{\infty} &= 
   \Bigg
   \{A(x)\in M(r,d) \Bigg|
   A(x)=
   \begin{pmatrix}
     0& {}^t\vec{w}_{d+1}\\ 
     \Vec{0}& O
   \end{pmatrix}
   x^{d+1}+
   \begin{pmatrix}
     v_d & {}^t\vec{w}_{d}\\
     \vec{0}&\tau
   \end{pmatrix}
   x^{d}
   +
   \begin{pmatrix}
     v_{d-1} & {}^t\vec{w}_{d-1}\\
     \vec{\nu}& T_{d-1}\\
   \end{pmatrix}
   x^{d-1}
    \\ 
    & \hspace{4cm} +\text{ lower terms in $x$}, 
     ~~~~T_{d-1} \in \mathcal{T}
    \Bigg\},
    \qquad \text{for $c = \infty$}.
\end{align*}
Here $\tau,\vec{\nu}$ and the set $\mathcal{T}$ is as follows:
\begin{align}\label{taunu}
  \begin{split}
  &\tau=
  \begin{pmatrix}
    0&0&\cdots &0\\
    1&0&\cdots&0\\    
    \vdots& \ddots &\ddots&\vdots\\
    0&\cdots&1&0
  \end{pmatrix}\
  \in M_{r-1}(\mathbb{C}),
  \qquad
  \vec{\nu}=\begin{pmatrix}1\\0\\\vdots\\0\end{pmatrix}
  \in\mathbb{C}^{r-1},
  \\
  &\mathcal{T} =
  \{ \rho \in M_{r-1}(\C) ~|~ \rho_{1j} = 0 \text{ for } j=1,\ldots,r-1 \}.
  \end{split}
\end{align}
By definition, 
$\mathcal{S}_c \subset \mathcal{M}_c$
since $\det D(A(x);x)=1$ for
all $A(x) \in \mathcal{S}_c$.
\begin{proposition}
  \label{decomp}
  For $c\in\mathbb{P}^1$, the map given by
  ${\mathcal S}_{c}\times G_r \to {\mathcal M}_c; ~(S(x),g(x))\mapsto g(S(x))$
  is an isomorphism. 
  Thus the space 
  ${\mathcal S}_{c}$ is a set of representatives of $\mathcal{M}_c / G_r.$
\end{proposition}
This is a consequence of the following lemma:

\begin{lemma}
  Let $c\in\mathbb{P}^1$.
  \begin{enumerate}
  \item
  If $A(x)\in {\mathcal M}_c$, then  
  there exists $g(x)\in G_r$ such that $g(A(x))\in {\mathcal S}_c$.
  \item
  If $g(S(x))= \tilde{S}(x)$ 
  with $S(x),\tilde{S}(x) \in {\mathcal S}_c$ and $g(x) \in G_r,$
  then we have $g(x)=\mathbb{I}_{r}$.
  \end{enumerate}
\end{lemma}

\begin{proof}
1:
We give a proof for $c \not= \infty.$
(The case of $c=\infty$ can be shown in a similar way.)
Define $B\in M_{r-1}(\mathbb{C})$ by
 \begin{equation}\notag
  B=\begin{pmatrix}
     \vec{u}(c),\zeta_1\vec{u}(c),\ldots, \zeta_{r-2}\vec{u}(c)
    \end{pmatrix}.
 \end{equation}
Here $\zeta_i$ $(1\leq i\leq r-2) \in M_{r-1}(\mathbb{C})$
are defined by
 \begin{equation}\notag
  \zeta_i=T(c)^i+\beta_1T(c)^{i-1}+\beta_2 T(c)^{i-2}+\cdots
  +\beta_i \mathbb{I}_{r-1},
 \end{equation}
where
$\beta_i$ $(1\leq i\leq r-1)$ 
are the coefficients of $y^i$ in the 
characteristic polynomial of $T(c)$:
$\det(y\mathbb{I}_{r-1}-T(c))=
y^{r-1}+\beta_1 y^{r-2}+\cdots+\beta_{r-1}$.
Since we have assumed $A(x)\in {\mathcal M}_c$,
$B$ is invertible.
Then we obtain
\begin{equation}\notag
\begin{pmatrix}
1&0\\0&B^{-1}
\end{pmatrix}
A(x)
\begin{pmatrix}
1&0\\0&B
\end{pmatrix}
=\begin{pmatrix}
*&*\\
\vec{\nu}&\ \tau'
\end{pmatrix}
+
(x-c)
\begin{pmatrix}
*&*\\ *& T
\end{pmatrix}
+\text{ higher terms in $(x-c)$ },
\end{equation}
where 
$$
  \tau' 
  = 
  \begin{pmatrix}
    -\beta_1 & -\beta_2 & \cdots & -\beta_{r-1}\\
    1 & 0 & \cdots& 0\\
    \vdots & \ddots & \ddots & \vdots\\
    0 & \cdots & 1 & 0
  \end{pmatrix}, 
  \qquad 
  T \in M_{r-1}(\C).
$$
We define $\vec{b}_1 $ and $\vec{b}_0$ by
\begin{equation}\notag
\vec{b}_1c+\vec{b}_0 = \,^t(\beta_1,\ldots,\beta_{r-1}),\qquad
\vec{b}_1=-{}^t(T_{11},T_{12},\ldots,T_{1 r-1}).
\end{equation}
Consequently we obtain the matrix 
\begin{equation}\notag
  g(x)
  =
  \begin{pmatrix} 1&0\\0&B \end{pmatrix}
  \begin{pmatrix} 1& ^t\vec{b}_1x+ \,^t\vec{b}_0\\0&1 \end{pmatrix},
\end{equation}
which satisfies $g(A(x))\in {\mathcal S}_c$.
\\
2: By expanding the relation $g(S(x))=\tilde{S}(x)$ in $(x-c)$
and comparing the coefficient matrices of $(x-c)^0$ and $(x-c)^1$,
we see $g(x)=\mathbb{I}_r.$
\end{proof}

\subsection{Integrable structure of $S_\infty$}

Now we set $c=\infty$.
We study an explicit relation between 
$\mathcal{S}_{\infty,P}$ and $\Div_{eff}^g(C_P)$,
then give a description of the vector field on $\mathcal{S}_{\infty}.$
These two results may be regarded as the counterparts of 
the studies on Beauville's system
by Smirnov and Zeitlin \cite{SmirnovZeitlin02} \S 4.1-2, and by
Fu \cite{Fu03} respectively.

Let $P \in V_{sm}(r,d)$ be such that
$\infty \in \mathbb{P}^1$ is not 
a ramification point of $\pi$,
and set $\mathcal{S}_{\infty,P} = \mathcal{S}_\infty \cap M_P$.
We study the relation between 
$\mathcal{S}_{\infty,P}$ and $\Div_{eff}^g(C_P)$ 
by applying the method of Sklyanin \cite{Sklyanin95}
(the separation of variables).
Let $\tau: \Div_{eff}^g(C_P) \to J_P^g$ be the Abel-Jacobi map.
Its restriction $\tau |_{\tau^{-1}(J_P')}$ is injective,
because the complete linear system of $L \in J_P '$ 
is of dimension zero (cf. Lemma \ref{degreeg}).
By abuse of notation, we write $\tau^{-1}$ for the composition of
\[  J_P' \overset{\cong}{\longrightarrow} \tau^{-1}(J_P')
    \hookrightarrow
    \Div_{eff}^g(C_P).
\]
Our aim is to give an explicit description of the composition $\kappa$ of
\[ \mathcal{S}_{\infty,P} \overset{\cong}{\longrightarrow}
   \mathcal{M_{\infty,P}}/G_r \overset{\cong}{\longrightarrow}
   J_P^g \setminus (\bigcup_{q\in \pi^{-1}(\infty)}\Theta_{q})
   \subset J_P'
   \overset{\tau^{-1}}{\longrightarrow}
    \Div_{eff}^g(C_P).
\]
Unfortunately, our result is limited to 
a subset of $\mathcal{S}_{\infty,P}$
due to technical difficulties.
Define
\[
{\mathcal S}_{\infty,P}'
   =\{A(x)\in {\mathcal S}_{\infty,P}|
  \text{ all roots of $ \det D(A(x);x)$ are simple 
            and belong to $\pi(C_P^0)$}\}.
\]
Note that
$\det D(A(x);x)$ of $A(x)\in{\mathcal S}_{\infty,P}$ 
is 
of degree $g$ by the definition of
${\mathcal S}_{\infty,P}$.

\begin{proposition}\label{prop:SoV}
Let $A(x)\in{\mathcal S}_{\infty,P}'$.
Denote by $x_1,\ldots,x_g$ the simple roots of $\det D(A(x);x)=0$.
Let $\vec{\nu} \in \mathbb{C}^{r-1}$ be any vector
satisfying
\begin{equation}\label{condition-nu}
\det(\vec{\nu},\vec{u}(x),\ldots, T(x)^{r-3}\vec{u}(x))
\not\equiv 0.
\end{equation}
With this $\vec{\nu}$,
define
\begin{equation}\label{separation-eigen}
  y_i:=\frac{
       \det(T(x)\vec{\nu},\vec{u}(x),\ldots, T(x)^{r-3}\vec{u}(x))}
       {\det(\vec{\nu},\vec{u}(x),\ldots, T(x)^{r-3}\vec{u}(x))}
       \Bigg|_{x=x_i}.
\end{equation}
(This is independent of the choice of $\vec{\nu}.$)
Then we have $\kappa(A(x)) = \sum_{i=1}^g(x_i,y_i).$
\end{proposition}

\begin{proof}
 The assumption that $x_i$ $(1\leq i\leq g)$
  is a simple root of $\det D(A(x);x)$
 implies that the rank of $D(A(x);x_i)$ is $r-2$.
 By Lemma \ref{lem:zero-entry},
 there exists a unique eigenvector of ${}^t A(x_i)$
 whose first component is zero.
 Denote the eigenvalue by $\alpha_i$.
 Then by Proposition \ref{prop:r-theta},
 the invertible sheaf
 $L$ corresponding to $A(x)$
 satisfies $L\in \bigcap_{i=1}^g \Theta_{(x_i,\alpha_i)}$.
 Because of the injectivity of 
 $\tau|_{\tau^{-1}(J_P')}$ mentioned above,
 we see $\kappa(A(x)) = \tau^{-1}(L) = \sum_{i=1}^g(x_i,\alpha_i).$
 Thus what we have to show is that $y_i=\alpha_i$.

 For simplicity, we show the case of $i=1$. 
 Since the eigenvalue $\alpha_1$ of $A(x_1)$ is also
 an eigenvalue of $T(x_1)$,
 there exists an eigenvector
 $\vec{\mu}'$ of $T(x_1)$
 of the eigenvalue $\alpha_1$.
 It is easy to show 
 $\det(\vec{\mu}'$, $\vec{u}(x_1)$,$\ldots$,$T(x_1)^{r-3}\vec{u}(x_1))
 \neq 0$, and we obtain
\begin{equation}\notag
\alpha_1=
 \frac{\det(T(x_1)\vec{\mu}',\vec{u}(x_1),\ldots,T(x_1)^{r-3}\vec{u}(x_1))}
      {\det(\vec{\mu}',\vec{u}(x_1),\ldots,T(x_1)^{r-3}\vec{u}(x_1))}.
\end{equation}
Let $\vec{\nu}\in {\mathbb{C}^{r-1}}$ be
a vector satisfying (\ref{condition-nu}).
Then there exist rational functions
  $\beta(x)$, $\beta_0(x)$,$\ldots$, $\beta_{r-3}(x)$
  $\in \mathbb{C}(x)$ 
such that
\begin{equation}\notag
   \vec{\nu}=
   \beta(x)\vec{\mu}'+\sum_{k=0}^{r-3}\beta_k(x)T(x)^{k}\vec{u}(x).
\end{equation}
Here 
    $\beta(x)\not\equiv 0$ 
by the assumption on $\vec{\nu}$.
Now it is immediate to check that  $y_1=\alpha_1$.
\end{proof}


Next we describe the vector field on ${\mathcal S}_\infty$
induced from $\eqref{Upsilon-field}$, using the following lemma:
\begin{lemma} 
  Let $X$ be a vector field on ${\mathcal M}_\infty \simeq 
  \mathcal{S}_\infty \times G_r$.
  The isomorphism 
  $\Phi: {\mathcal M}_{\infty} \stackrel{\sim}{\to} 
  {\mathcal S}_{\infty}\times G_r;
  ~ A(x) \mapsto (S(x),g(x))$ induces the decomposition of $X$
  as $\Phi_*X = F + G$,
  where 
  $F\in H^0({\mathcal S}_{\infty}\times G_r, T{\mathcal S}_{\infty})$
  and 
  $G\in H^0({\mathcal S}_{\infty}\times G_r, TG_r)$.
  Then
  \begin{equation} \label{proj-Sinfty}
    X(A(x)) = g\bigl(F(S(x),g(x))\bigr) - [g(x)^{-1}G(S(x),g(x)), A(x)].
  \end{equation}
  Here we identify $T_{g(x)}G_r$ with $Lie G_r$, and 
  $T_{S(x)}S_\infty$ with the subspace of $M(r,d)$
  via the inclusion $S_\infty \hookrightarrow M(r,d)$.
 \end{lemma}

The proof is left to the reader.
The Hamiltonian vector field on $\mathcal{S}_\infty$ becomes as follows:
\begin{proposition}
  \label{prop:Y-projection}
  The projection of the vector field \eqref{Upsilon-field}
  onto ${\mathcal S}_{\infty}$ is
  \begin{equation}\label{Svector-field} 
    F^{(p)}_a(A(x)) = 
    \frac{1}{x-a}[A(a)^p,A(x)]+
    \Bigg[
    \begin{pmatrix}0& ^t\vec{\gamma}_p x+ \,^t\vec{\beta}_p\\
    \vec{0}&C_p\end{pmatrix},A(x)
    \Bigg]
    ~~~~
    \text{ at }
    A(x)\in {\mathcal S}_{\infty}.
  \end{equation}
  Here $(\vec{\gamma}_p,\vec{\beta}_p,C_p) \in 
  \mathbb{C}^{r-1} \oplus \mathbb{C}^{r-1} \oplus M_{r-1}(\mathbb{C})$ 
  is a unique solution of
  \begin{equation}\label{proj}
    \begin{split}
    &C_p\cdot \vec{\nu}=(\tau-v_d \mathbb{I}_{r-1})\cdot \vec{h}_p,
    \\
    &\vec{\nu}\cdot \,^t\vec{\gamma}_p-[C_p,\tau]
    =\vec{h}_p\cdot \,^t\vec{w}_{d+1},
    \\
    &(\vec{\nu}\cdot \,^t\vec{\beta}_p
    +\vec{u}_{d-2}\cdot \,^t\vec{\gamma}_p- C_p T_{d-1})_{1,i}
    \\
    & ~~~=
    (\vec{h}_p\cdot(^t\vec{w}_d+a \,^t\vec{w}_{d+1})+J_p \tau)_{1,i},
    \text{ for $1 \leq i \leq r-1$}, 
    \end{split}
  \end{equation}
  where
  $\tau$ and $\vec{\nu}$ are defined in (\ref{taunu}), 
  and $\vec{h}_p$ and $J_p$ are
  \begin{equation}
    \label{A^p}
    A(a)^p=\begin{pmatrix}
    *&*\\\vec{h}_p & J_p
    \end{pmatrix}.
  \end{equation}
\end{proposition}
\begin{proof}
The equations \eqref{proj} are obtained by solving 
\eqref{proj-Sinfty} for 
$F(S(x),g(x))$ and $G(S(x),g(x))$ 
at $X = \Upsilon_a^{(p)}$ and
$g={\mathbb{I}_r}$.
Eq. \eqref{proj-Sinfty} becomes
\begin{equation}\label{FG}
   \Upsilon_a^{(p)}(A(x))
   =
   F_a^{(p)}(A(x)) - [ G_a^{(p)}(A(x)) \,,\,A(x)], 
\end{equation}
where $F_a^{(p)}(A(x))$ is of the form
\begin{equation*}
  F_a^{(p)}(A(x))=
  x^{d+1}
  \begin{pmatrix}
    0&*\\\Vec{0}&O
  \end{pmatrix}
  +x^d
  \begin{pmatrix}
    *&*\\\Vec{0}&O
  \end{pmatrix}
  +x^{d-1}
  \begin{pmatrix}
    *&*\\\Vec{0}&\rho
  \end{pmatrix}
  +
  \text{ lower terms in $x$}.
\end{equation*}
Here $\rho \in \mathcal{T}$ \eqref{taunu},
and $G_a^{(p)}(A(x)) \in \mathrm{Lie} \,G_r$ 
is of the form
\begin{equation*}
  G_a^{(p)}(A(x)) 
  =
  \begin{pmatrix}
    0& ^t\vec{\gamma}_p x+ \,^t\vec{\beta}_p\\
    \vec{0}&C_p
  \end{pmatrix}
  \qquad
  (\vec{\gamma}_p,\vec{\beta}_p\in \mathbb{C}^{r-1},
   C_p\in M_{r-1}(\mathbb{C})
  ).
\end{equation*}
The matrix $G_a^{(p)}(A(x))$ is determined 
as follows.
In the LHS of \eqref{FG}, the $(i,1)$-entries $(2\leq i\leq r)$ and
$(i,j)$-entries $(2\leq i,j\leq r)$ are
\begin{equation}\notag
  \begin{split}
  &
  (v_d \mathbb{I}_{r-1}-\tau)\cdot \vec{h}_p \,x^{d-1}
    +\text{ lower order in $x$},
  \\
  &
  \vec{h}_p\cdot \, ^t\vec{w}_{d+1} \,x^d+
  \big(
  \vec{h}_p\cdot ( \, ^t\vec{w}_d+a \,^t\vec{w}_{d+1})+[J_p,\tau]\big)x^{d-1}
  +\text{ lower order in $x$}.
  \end{split}
\end{equation}
In the RHS of \eqref{FG},the $(i,1)$-entries $(2\leq i\leq r)$ and
$(i,j)$-entries $(2\leq i,j\leq r)$ are
\begin{equation}\notag
  \begin{split}
  &
  -C_p\cdot \vec{\nu}\,x^{d-1}+\cdots,
  \\ 
  &
  (\vec{\nu}\cdot \, ^t\vec{\gamma}_p-[C_p,\tau])x^d+
  (\vec{\nu}\cdot \, ^t\vec{\beta}_p+\vec{u}_{d-2}\cdot \, ^t\vec{ \gamma}_p
  -[C_p,T_{d-1}]+
  \rho
  )x^{d-1}+\cdots.
  \end{split}
\end{equation}
We obtain the equations \eqref{proj}
for $(\vec{\gamma}_p,\vec{\beta}_p,C_p)$
by comparing the LHS and the RHS. 

The solution to eqs.~\eqref{proj} is unique
since
the first and second equations 
completely determine $C_p$ and $\vec{\gamma}_p$
and then the third equation 
completely determines the value of $\vec{\beta}_p$.
\end{proof}

\subsection{Examples}
 
The case of $r=2$:
we have the space of representatives as
\begin{align*}
  \mathcal{S}_\infty = 
  \Bigg\{
  A(x)&= \begin{pmatrix}
          v(x) & w(x) \\
          u(x) & t(x) 
          \end{pmatrix}
        \\
      &=
        \begin{pmatrix}
          0 & w_{d+1} \\ 
          0 & 0
        \end{pmatrix} x^{d+1}
        +
  \begin{pmatrix}
    v_d & w_d \\
    0 & 0
  \end{pmatrix} x^{d}
  +
  \begin{pmatrix}
    v_{d-1} & w_{d-1}\\
    1 & 0 \\
  \end{pmatrix} x^{d-1}
  +\text{ lower terms in $x$}
  \Bigg\}.
\end{align*}
For $P \in V_{sm}(r,d)$, the genus of the curve $C_P$ is $d-1$.
The isomorphism given in Proposition \ref{prop:SoV}
becomes very simple:
$x_k ~(k=1, \cdots, d-1)$ are the zeros 
of $u(x)$ and $y_k = t(x_k)$.  
The vector field on $\mathcal{S}_\infty$ \eqref{Svector-field} 
becomes
  \begin{equation}\notag
    F_a^{(1)}(A(x)) 
    =
    \Bigg[
      \frac{1}{x-a} A(a) 
      +
      u(a) 
      \begin{pmatrix}
        0&(x+a-u_{d-2})w_{d+1}+w_d\\
        0&-v_d
      \end{pmatrix}
      ~,~
      A(x)
     \Bigg].
  \end{equation}
Let $V = \{ P(x,y) \in V(2,d) ~|~ s_1(x) \equiv 0 \}.$
The restriction $\psi^{-1}(V) \cap \mathcal{S}_\infty \to V$
of our system
$\psi |_{\mathcal{S}_\infty}: \mathcal{S}_{\infty} \to V(2, d)$
coincides with the even Mumford system 
introduced by Vanhaeche \cite{Van1638}.

The case of $r=3$: this is a new system.
$\mathcal{S}_\infty$ is written as
\begin{align*}
  \mathcal{S}_\infty
  =
  \Bigg\{
  A(x)&= \begin{pmatrix}
          v(x) & w^{(1)}(x) & w^{(2)}(x) \\
          u^{(1)}(x) & T^{(1,1)}(z) & T^{(1,2)}(z) \\
          u^{(2)}(x) & T^{(2,1)}(z) & T^{(2,2)}(z) 
          \end{pmatrix}
        \\
      &=
        \begin{pmatrix}
          0 & w_{d+1}^{(1)} & w_{d+1}^{(2)}\\ 
          0 & 0 & 0 \\
          0 & 0 & 0
        \end{pmatrix} x^{d+1}
        +
  \begin{pmatrix}
    v_d & w_{d}^{(1)} & w_{d}^{(2)}\\ 
    0 & 0 & 0 \\
    0 & 1 & 0
  \end{pmatrix} x^{d}
  +
  \begin{pmatrix}
    v_{d-1} & w_{d-1}^{(1)} & w_{d-1}^{(2)}\\ 
    1 & 0 & 0 \\
    0 & T_{d-1}^{(2,1)} & T_{d-1}^{(2,2)} 
  \end{pmatrix} x^{d-1}
  \\
  & \qquad
  +\text{ lower terms in $x$}
  \Bigg\}.
\end{align*}
For $P \in V_{sm}(r,d)$, the genus $g$ of $C_P$ is $3d-2$.
The isomorphism given in Proposition \ref{prop:SoV} becomes as follows:
$x_k$ are the zeros of $D(A(x))$ \eqref{D-function} 
and $y_k$ \eqref{separation-eigen} has two equivalent descriptions:
$$
  y_k = \frac{u^{(2)}(x) T^{(1,1)}(x) - u^{(1)}(x) T^{(2,1)}(x)}
             {u^{(2)}(x)}
        \Big|_{x= x_k} 
        \text{ or }
        \frac{u^{(1)}(x) T^{(2,2)}(x) - u^{(2)}(x) T^{(1,2)}(x)}
             {u^{(1)}(x)}
        \Big|_{x= x_k}. 
$$  
The vector field on $\mathcal{S}_\infty$ is written as 
  \begin{equation}\notag
    F_a^{(p)}(A(x)) 
    =
    \Bigg[
      \frac{1}{x-a} A(a) 
      +
      \begin{pmatrix}
        0 & ^t\vec{\gamma}_p x + \, ^t\vec{\beta}_p \\ 
        0 & C_p 
      \end{pmatrix}
      ~,~
      A(x)
     \Bigg]
   \qquad \text{ for } p=1,2,
  \end{equation}
where
\begin{align*}
  ^t\vec{\gamma}_p x +\,^t\vec{\beta}_p 
  &=
  h^{(1)}_p \bigl( (x + a - u^{(1)}_{d-2}) \,^t\vec{w}_{d+1} + \,^t\vec{w}_d
            \bigr)
  \\
  & \qquad + \,^t\bigl(h^{(2)}_p w_{d+1}^{(2)}(x+T_{d-1}^{(2,1)}-u_{d-2}^{(1)})
                   + (J_p)_{1,2} ~,~
             h^{(2)}_p w_{d+1}^{(2)} T_{d-1}^{(2,2)}
             \bigr)
  \\
  C_p 
  &=
  h^{(1)}_p
  \begin{pmatrix}
    -v_d & 0 \\
    1 & -v_d  
  \end{pmatrix}
  +
  h^{(2)}_p
  \begin{pmatrix}
    1 & w_{d+1}^{(2)}\\
    -v_d & -w_{d+1}^{(1)}
  \end{pmatrix}.
\end{align*}
Here $J_p$ and 
$\vec{h}_p = 
  \begin{pmatrix}
    h^{(1)}_p \\
    h^{(2)}_p 
  \end{pmatrix}$
are given at \eqref{A^p}.


\end{document}